\documentclass[12pt, peerreview , onecolumn,letterpaper]{IEEEtran}

\pdfoutput=1

\RequirePackage[normalem]{ulem} 
\RequirePackage{color}\definecolor{RED}{rgb}{1,0,0}\definecolor{BLUE}{rgb}{0,0,1} 

\RequirePackage{color}

\usepackage{tikz}
\usepackage[sumlimits,intlimits]{amsmath}
\usepackage{amsmath,amsfonts,amssymb}%
\usepackage{bm}%
\usepackage{graphicx,graphics}%
\usepackage{cases}%
\usepackage[noadjust]{cite}%
\usepackage{color}%
\usepackage{cite,url}
\usepackage{verbatim}
\usepackage{enumerate}
\usepackage{algorithm}
\usepackage{balance}
\usepackage{multirow}
\usepackage{stfloats}
\usepackage{subfigure}
\usepackage{xspace}
\usepackage{wrapfig}
\usepackage{smartref}

\ifCLASSOPTIONpeerreview

\fi

\begin{document}

\title{Hybrid Millimeter-Wave Systems: A Novel Paradigm for HetNets}

\author{\vspace*{-8pt}
\authorblockN{Hani Mehrpouyan,~\IEEEmembership{Member, IEEE},
                Michail Matthaiou,~\IEEEmembership{Senior Member, IEEE},
                Rui Wang,~\IEEEmembership{Member, IEEE},
                George K. Karagiannidis,~\IEEEmembership{Fellow, IEEE,}
                and
                Yingbo Hua,~\IEEEmembership{Fellow, IEEE}
              }\\
}
\maketitle
\thispagestyle{empty}
{\let\thefootnote\relax\footnotetext{{Hani Mehrpouyan is with the Department of Electrical and Computer Engineering, California State University. Michail Matthaiou is with the School of Electronics, Electrical Engineering and Computer Science, Queen's University Belfast, U.K., and with the Department of Signals and Systems, Chalmers University of Technology, Sweden. Rui Wang is with Institute of Network Coding, The Chinese University of Hong Kong. George K. Karagiannidis is with the Department of Electrical and Computer Engineering, Aristotle University of Thessaloniki, Greece. Yingbo Hua is with the Department of Electrical Engineering at the University of California, Riverside. Emails: hani.mehr@ieee.org, m.matthaiou@qub.ac.uk, liouxingrui@gmail.com, geokarag@ieee.org, and yhua@ee.ucr.edu.
\vspace{-0pt}}} } 

\begin{abstract}

\vspace*{-0pt}

Heterogeneous Networks (HetNets) are known to enhance the bandwidth efficiency and throughput of wireless networks by more effectively utilizing the network resources. However, the higher density of users and access points in HetNets introduces significant inter-user interference that needs to be mitigated through complex and sophisticated interference cancellation schemes. Moreover, due to significant channel attenuation and presence of hardware impairments, e.g., phase noise and amplifier nonlinearities, the vast bandwidth in the millimeter-wave band has not been fully utilized to date. In order to enable the development of multi-Gigabit per second wireless networks, we introduce a novel millimeter-wave HetNet paradigm, termed \emph{hybrid} HetNet, which exploits the vast bandwidth and propagation characteristics in the $60$ GHz and $70$--$80$ GHz bands to reduce the impact of interference in HetNets. Simulation results are presented to illustrate the performance advantage of hybrid HetNets with respect to traditional networks. Next, two specific transceiver structures that enable hand-offs from the   $60$ GHz band, i.e., the \emph{V-band} to the $70$--$80$ GHz band, i.e., the \emph{E-band}, and vice versa are proposed. Finally, the practical and regulatory challenges for establishing a hybrid HetNet are outlined.

\end{abstract}

\newpage

\vspace*{-8pt}
\section{Introduction}

\vspace{+3pt}
Current generation of wireless standards, i.e., IEEE $802.11$ac, can support data rates of up to $1.6$ Gigabit per second (Gbps) while employing high-order modulations and multiple-input and multiple-output technology. In comparison, the latest wired Ethernet protocol can easily support data rates of up to $100$ Gbps, i.e., more than 60 times faster. Thus, to ensure that wireless networking continues to be a viable alternative to wired networks, there is an urgent need for the development of multi-Gbps wireless links.

\begin{wrapfigure}{r}{0.63\textwidth}
\begin{center}
\vspace*{-20pt}
\hspace*{-5pt} \scalebox{0.54}{\includegraphics {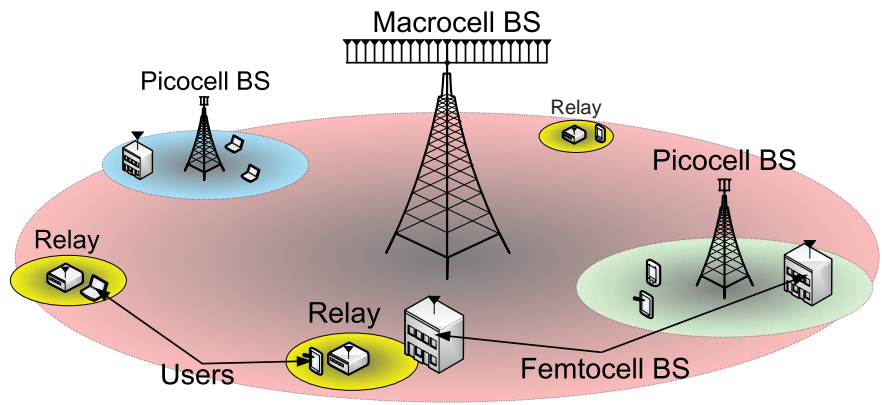}}
\vspace*{-300pt}
\caption{A HetNet with a macrocell BS and multiple supporting picocell BSs, femtocells, and relays.}
\vspace*{-10pt}
\label{fig:HET_SYSTEM_SETUP}
\end{center}
\end{wrapfigure}

\vspace{+3pt}
One approach for meeting the above need has been to more effectively share the network resources through the widely accepted notion of \emph{heterogeneous networks (HetNets)}, see Fig.~\ref{fig:HET_SYSTEM_SETUP}. By employing smaller and more specialized cells, such networks can more efficiently meet the users' needs and improve the overall throughput of cellular networks \cite{article_andrews13, article_Ghosh12, Hoymann-12-A, book_FEMTO_IMP, article_metis, article_hetnet_synch}.

However, the close vicinity of many users and cellular base station (BSs) along with the interference amongst these devices in HetNets, have introduced new challenges to the design of communication systems. Although many algorithms and approaches have been proposed for interference management and alignment in HetNets \cite{article_Barbieri12}, these schemes are mainly complex in nature and may not be suitable for cost and power sensitive wireless applications.

\vspace{+3pt}
Another approach for meeting the increasing demands for faster data connectivity is to take advantage of the vast bandwidth in the millimeter-wave band. It is noteworthy that the commissioned bandwidth by the Federal Communication Commission (FCC) in the $60$ and $70$--$80$ GHz bands alone is fifty times the available bandwidth in today's cellular networks \cite{fcc_09}. Thus, the millimeter-wave spectrum provides a great potential for meeting the tremendous demand for affordable and ultra high-speed wireless links. Current research has shown that point-to-point systems in the $70$--$80$ GHz or the \emph{E-band} can indeed support significantly higher date rates than the systems using the microwave band (frequencies below $30$ GHz) \cite{book_Georgiadis-13}. Moreover, new wireless networking standards, e.g., IEEE $802.11$ad and IEEE $802.15.3$c, are designed to support multi Giga bit per second (Gbps) wireless networks in the $60$ GHz or the \emph{V-band}. In spite of these positive developments, millimeter-wave systems are significantly less bandwidth efficient than their counterparts in the microwave band \cite{Mehrpouyan-14-M,Dyadyuk-07-J,Cheng_11,wells_10}.

\begin{wrapfigure}{r}{0.6\textwidth}
\begin{center}
\vspace{-45pt}
\hspace*{-5pt}\scalebox{0.7}{\includegraphics {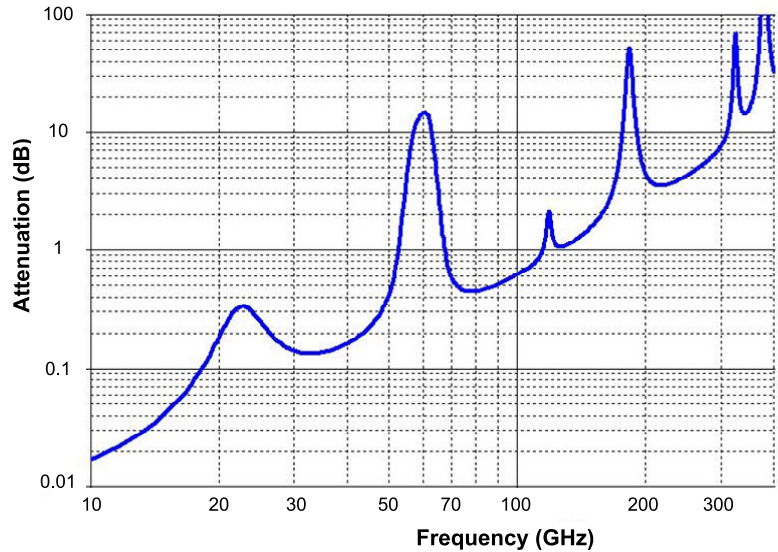}}
\vspace*{-9.0cm}
\caption{Atmospheric attenuation vs. operating frequency \cite{wells_10}.}
\label{fig:attenuation}
\end{center}
\vspace{-10pt}
\end{wrapfigure}

\vspace{+3pt}
As shown in Fig.~\ref{fig:attenuation}, compared to the $60$ GHz band, the atmospheric absorption in the $70$--$80$ GHz spectrum is much lower, i.e., approximately $16$ dB lower. Moreover, the FCC regulations allow for higher transmission power in the E-band compared to the V-band, i.e., maximum transmit power of $0.5$ W and $3$ W for the V- and E-bands, respectively \cite{wells_10}. Accordingly, due to the large attenuation factor and high antenna directivity, the $7$ GHz of bandwidth in the $60$ GHz band can be used to establish ultra high-speed wireless links without causing significant interference to neighboring devices and networks. These characteristics, which are beneficial from an interference management point-of-view, can also limit the ability of V-band systems to meet the quality of service (QoS) and throughput requirements of users in wireless or cellular networks.


%
%
%

\vspace{+3pt}
Using the aforementioned characteristics of the V- and E-bands, in this work, we present a novel hybrid millimeter-wave based HetNet paradigm that enhances the bandwidth efficiency of millimeter-wave systems while reducing interference. More specifically, a hybrid HetNet makes use of the large bandwidth in the V-band to establish short range ultra high-speed point-to-point links. Due to the strong radio signal attenuation and high antenna directivity in the V-band, such links will not be a significant source of interference. Moreover, to circumvent the shortcomings of the V-band and to meet the QoS requirements of the network, we propose to apply the E-band spectrum to establish the longer range links and to interconnect the HetNet BSs. For example, the links between the macro- and picocell BSs and users can operate in the E-band, while the V-band can be used by femtocell BSs. Our simulation results illustrate the significant advantages of utilizing both the V- and E-bands in hybrid HetNet topologies. Next, new transceiver structures that allow for seamless handover from the V- to E-band and vice versa are presented and their advantages and disadvantages are discussed. Finally, the challenges of implementing a hybrid HetNet structure in the millimeter-wave band, due to regulatory limitations, are presented and preliminary solutions are proposed.

\vspace{+3pt}
The remainder of this article is organized as follows: In Section \ref{sec:framework}, the potential of utilizing the V- or/and E-bands in hybrid HetNets is presented. Section \ref{sec:handoff} focuses on the transceiver structures and hand-off issues in hybrid HetNets. Section \ref{sec:implementation} outlines the main regulatory challenges for developing and implementing a Hybrid HetNet structure and presents preliminary solutions to these challenges.


\vspace*{-8pt}

\section{Potential of Hybrid HetNets}\label{sec:framework}

The main challenges for the wide deployment of next generation wireless networks in the V-band can be summarized as \cite{book_Georgiadis-13,wells_10,Cheng_11}: 1) high path loss; 2) significant signal attenuation due to shadowing; and 3) limited transmit power. There are a number of solutions that are proposed for tackling these issues. These range from the deployment of highly directional antennas \cite{article_yong07,Cheng_11,article_artemenko13,book_Hansen-98} to the application of multi-beam directional arrays that radiate in multiple directions \cite{article_lee_12-gsmm, article_Huang_06-3}. The former approach requires the application of tracking algorithms that estimate the location of a user and adjust the beam pattern of the antenna mechanically or through signal processing approaches. Hence, this approach can be complex to implement and its performance is highly dependent on the accuracy of the tracking algorithm, which can vary substantially based on the propagation environment. By transmitting in multiple directions, the second approach seeks to mitigate the shadowing issue in the V-band, since it is anticipated that at least one of the beams or its reflections will reach the receiver. However, none of these schemes can address the limited operational range of V-band systems due to the significant channel attenuation and oxygen absorption.

\vspace{+3pt}
Unlike the V-band, the available spectrum in the E-band does not suffer from significant channel attenuation. As such, the $10$ GHz of bandwidth in the E-band can be utilized to establish links with longer operational ranges \cite{wells_10}. Although these characteristics are advantageous and can address the shortcomings of the V-band, they can result in significant interference in densely deployed HetNets. Thus we propose to concurrently use the bandwidths in the V- and E-bands to achieve higher data rates, while reducing interference. Table \ref{table_1} summarizes the framework for utilizing each or both bands in a hybrid HetNet configuration based on the network density and link SNR. According to Table \ref{table_1}, a hybrid HetNet topology can use the V-band in densely deployed networks. In this scenario, the high channel attenuation in the V-band ensures that the overall level of interference is reduced in the network. In addition, to maintain the QoS in low signal-to-noise ratio (SNR) scenarios, the E-band spectrum can be used instead. The lower channel attenuation and higher transmission power in the E-band can be used to increase the SNR. Finally, when extremely high throughputs are needed and the interference levels are low, both bands can be utilized.

\begin{table}[t]
  \centering
  \caption{Hybrid Hetnet Allocation of the V- and E-band Spectra}\label{table_1}
  \begin{tabular}{ |c | c | c |}
  \hline
  \emph{Low SNR} &	\emph{Medium-High SNR} & \emph{Medium-High SNR}\vspace{-7pt}\\
  \& &	\& & \&\vspace{-7pt}\\
  \emph{Low/High Density} &	\emph{High Density} & \emph{Low Density}\\
  \hline
  \hline
  E-band&V-band& both V- and E-bands\\
  \hline
\end{tabular}
\end{table}

\begin{wrapfigure}{r}{0.60\textwidth}
\begin{center}
\vspace*{-10pt}
\hspace*{-8pt}\scalebox{0.55}{\includegraphics {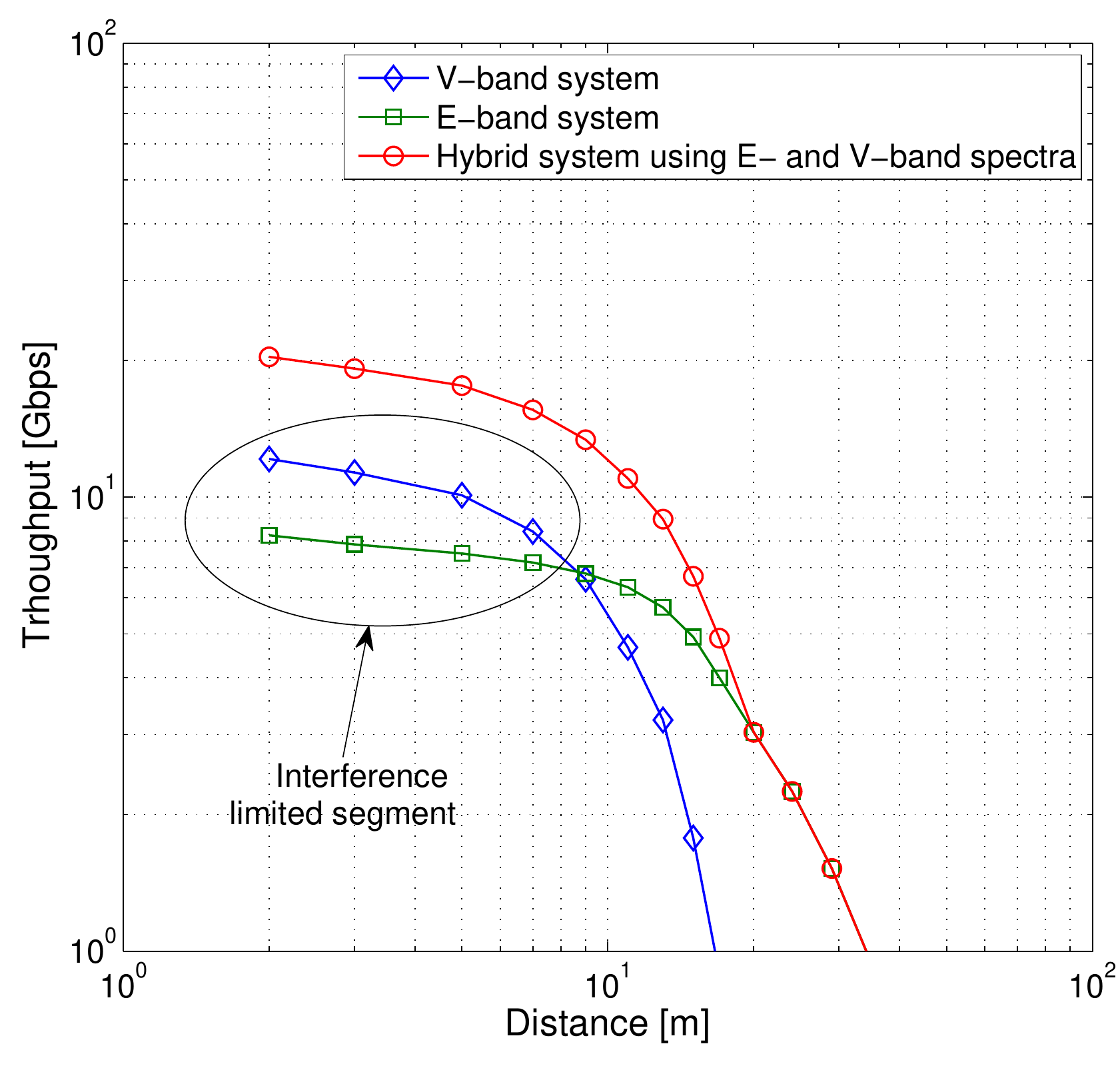}}
\vspace*{-20pt}
\caption{Throughput of a HetNet system when operating in either the V- or E-band, or applying \emph{both} spectra (combined antenna gain of $30$ dBi, bandwidth $5$ GHz for both the V- and E-band, and 10-dB human shadowing).}
\vspace*{25pt}
\label{fig:hybrid_hetnet}
\end{center}
\end{wrapfigure}

\vspace{+3pt}
In order to demonstrate the potential of a hybrid HetNet configuration, in Fig.~\ref{fig:hybrid_hetnet}, we compare the throughput of a point-to-point wireless link, while considering systems operating in the V-band, the E-band, and both V- and E-bands simultaneously. In this scenario, the effects of channel attenuation, human shadowing, and interference from neighboring devices on the throughput of the link are also taken into account. Two tiers are taken into consideration. The network is assumed to consist of a macrocell and a femtocell BS and ten users. The x-axis indicates the average distance amongst the users and their respective BSs. Our numerical results in Fig.~\ref{fig:hybrid_hetnet} demonstrate the advantage of a hybrid HetNet compared to a HetNet that is operating in either the V- or E-band. As shown in Fig.~\ref{fig:hybrid_hetnet}, as the link distance increases, the

\vspace{-25pt}
\noindent
throughput of an ``E-band system" is significantly higher than that of a ``V-band system" due to a lower channel attenuation factor and higher maximum regulated transmission power. However, Fig.~\ref{fig:hybrid_hetnet} shows that when taking the effect of interference into account, as the network density increases, a V-band wireless system can outperform its E-band counterpart. This can be attributed to the higher attenuation in the V-band, which reduces the overall inter-user interference in the network. As such, the results in Fig.~\ref{fig:hybrid_hetnet} depict that a HetNet that can take advantage of \emph{either} the V- or E-band can support wireless links over longer distances while reducing interference in densely deployed networks. Moreover, we observe that the overall performance of a wireless network can be further improved if the transceivers used in the hybrid HetNet can \emph{simultaneously} transmit information over both the V- and E-band spectra. This result is denoted by ``Hybrid system using E- and V-band spectra" in Fig.~\ref{fig:hybrid_hetnet}.

Based on the results in Fig.~\ref{fig:hybrid_hetnet}, the precursor for transitioning or handing-off from one band to another can be due to the following reasons:
\begin{itemize}
\item In the strong interference regime, the access point or BS serving the wireless or cellular network, respectively allocates the V-band spectrum to users with high SNRs, while assigning the E-band spectrum to users at lower SNR values. For example, the femtocells within a hybrid HetNet may use the V-band spectrum to avoid interfering with neighboring devices that are communicating with the macrocell or picocell BSs.
\item In the low SNR scenario, to maintain the QoS, the access point allocates the E-band spectrum to users that have a link SNR below a certain threshold. For example, the E-band can be used to to establish longer range links that connect users at the cell edges to the macrocell BS. 

\item In the high SNR and low interference scenario, to meet the higher throughput requirements of a user, the access points may utilize the spectrum in both the V- and E-bands to  meet the users' demands. This scenario can be considered in the context of transmitting audio/visual signals to an entrainment center. Such high-speed links can also be used to interconnect BSs.
\end{itemize}


\vspace*{-12pt}
\section{Transceiver Structures for Handover Between V- and E-band}\label{sec:handoff}
\begin{wrapfigure}{r}{0.68\textwidth}
\begin{center}
\vspace*{-20pt}
\hspace*{25pt}\scalebox{0.65}{\includegraphics {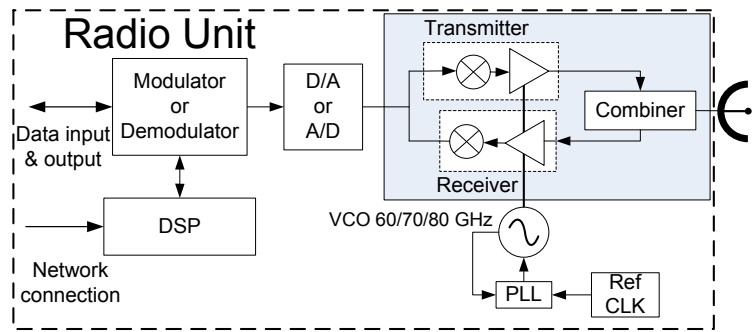}}
\vspace*{-320pt}
\caption{Transceiver structure for utilizing either the V- or E-band.}
\vspace*{-10pt}
\label{fig:simple_hetnet}
\end{center}
\end{wrapfigure}

\vspace{+3pt}
Since the process of transitioning from the V- to E-band in hybrid HetNets is directly influenced by the transceiver design, we first propose two specific transceiver structures that
\begin{wrapfigure}{r}{0.68\textwidth}
\begin{center}
\vspace*{-25pt}
\hspace*{10pt}\scalebox{0.53}{\includegraphics {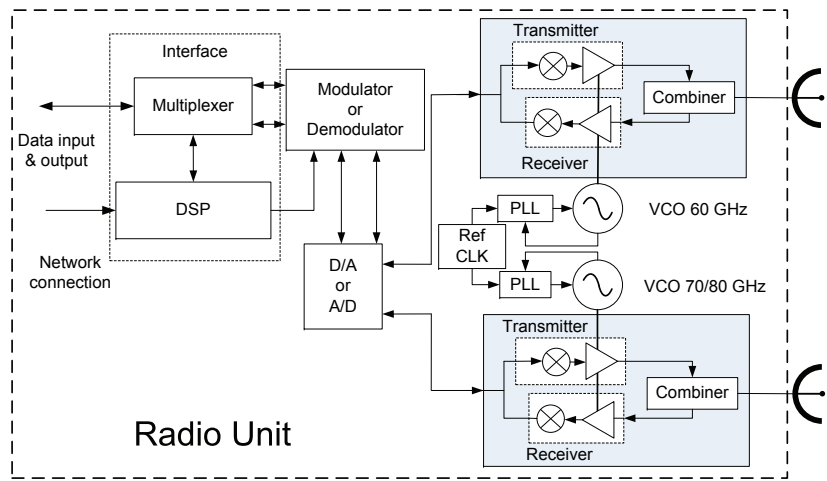}}
\vspace*{-180pt}
\caption{Transceiver structure for utilizing both the V- and E-band spectra simultaneously (PLL, DSP, and CLK stand for phase locked loop, digital signal processing, and clock, respectively).}
\vspace*{-15pt}
\label{fig:complex_hetnet}
\end{center}
\end{wrapfigure}

\noindent  support communication links in both the V- and/or E-band spectra. The transceiver design in Fig.~\ref{fig:simple_hetnet} allows for signal transmission in either the V- or E-band, while the transceiver structure  in Fig.~\ref{fig:complex_hetnet} supports simultaneous transmission in both bands. The advantages and disadvantages of each design can be summarized as follows:

\begin{itemize}
\item Since the transceiver design in Fig.~\ref{fig:simple_hetnet} uses a single band at any given time, it is possible to use a single voltage controlled oscillator (VCO) and amplifier at the radio frequency (RF) end. Moreover, as shown in Fig.~\ref{fig:simple_hetnet}, this transceiver structure does not require the deployment of a multiplexing block for transmission of the information bits in both the V- and E-bands. Therefore, the transceiver design in Fig.~\ref{fig:simple_hetnet} is less complex and less costly to implement when compared to the design in Fig.~\ref{fig:complex_hetnet}.

\item In general, designing an accurate and cost effective VCO that can operate over a very large set of frequencies is a rather challenging task. This design constraint combined with the fact that the transceiver in Fig.~\ref{fig:simple_hetnet} uses a single oscillator for $60$, $70$, and $80$ GHz band transmission, may make this transceiver more susceptible to the negative impact of disturbances, such as oscillator phase noise, compared to the transceiver in Fig.~\ref{fig:complex_hetnet}.
    %
%
%
%
%
%

\item One of the main challenges in the area of millimeter-wave communications is the design of RF amplifiers that can efficiently operate in this band. Keeping this in mind we note that the transceiver design in Fig.~\ref{fig:simple_hetnet} requires the application of an amplifier that can operate over a span of $30$ GHz. Therefore, considering today's technology, it is anticipated that the transceiver in Fig.~\ref{fig:simple_hetnet} may be more severely affected by amplifier nonlinearities and limited transmit power. On the other hand, the design in Fig.~\ref{fig:complex_hetnet}, may deploy standard amplifiers that are designed for either the V- or E-band. This extensively simplifies the design process and may reduce the impact of impairments such as amplifier nonlinearities.

\item The transceiver design in Fig.~\ref{fig:complex_hetnet} can support higher data rates than the approach in Fig.~\ref{fig:simple_hetnet} since it can use the spectra in both the V- and E-bands simultaneously.

\end{itemize}

Seamless handover between the V- and E-band communication is essential for the effective operation of hybrid HetNets.~Although, very sophisticated handover schemes are available for current microwave based cellular networks \cite{article_kim_10,book_parkvall-11}, these schemes may not be straightforwardly applied in the case of millimeter-wave systems, where the handovers take place over a much wider range of frequencies. For example, since the center frequencies for the V- and E-band are separated by as much as $10$ GHz, the transceivers intended for use in hybrid HetNets may require the application of two separate oscillators for each band. As such, compared to the microwave band, the handover between millimeter-wave frequencies is expected to be more severely affected by impairments, such as frequency offset and oscillator phase noise. This renders the process of achieving carrier synchronization during the handover process between the V- and E-bands to be far more challenging than microwave based cellular networks. In addition, the handover schemes proposed for current cellular networks are designed to support transition from one cellular BS to another \cite{article_kim_10,book_parkvall-11}, while in hybrid HetNets the transition from the V- to E-band communication takes place
 amongst the same set of transceivers. This specific characteristic can be applied to reduce the overhead and complexity associated with handovers in hybrid HetNets. For instance, in the transceiver design in Fig.~\ref{fig:complex_hetnet}, while handing off from one band to another, the current link can serve as a feedback link to ease the process of channel estimation and synchronization during the handoff process. Thus, further research is needed to develop a comprehensive approach for seamless transmission from the V- to E-band and vice versa in hybrid HetNet configurations.

%
%
%

\vspace*{-8pt}
\section{Regulatory Issues for Development of Hybrid HetNets}\label{sec:implementation}
Although the previous sections have demonstrated the advantages of using a hybrid HetNet structure, there are some regulatory issues that need to be addressed to make the deployment of such networks possible. In this section, we briefly summarize these issues and provide solutions to each specific case.

The V- and E-band spectra are regulated or are being considered for regulation for the deployment of communication systems by most countries and regions in the world. More specifically, the V-band spectrum has been regulated for use in the United States, European Union, China, Canada, Korea and many other countries around the world. Moreover, the E-band spectrum is regulated both in the United States and Europe and is being considered for the deployment of wireless communication systems in China, Canada, and many other countries. Although these regulations somewhat vary from country to country, the regulations proposed by FCC represent a good sample of what is proposed or being proposed for these bands. We now summarize the FCC regulations regarding the use of both bands in Table~\ref{table_2}.
\begin{table}[t]
  \centering
  \caption{FCC Regulatory Rules for Utilizing the V- and E-bands \cite{wells_10}}\label{table_2}
  \begin{tabular}{|l |c | c |}
  \hline
  & \emph{V-band} &	\emph{E-band}\\
  \hline
  \hline
  Frequency Range &$57$--$64$ GHz& $71$--$76$, $81$--$86$, $92$--$95$ GHz\\
  \hline
  Licensing &Unlicensed& Licensed\\
  \hline
  Maximum Transmit Power &$27$ dBm& $35$ dBm\\
  \hline
  Minimum Antenna Gain & Not Applicable& $43$ dBi\\
  \hline
\end{tabular}
\end{table}

Based on the regulatory requirements presented in Table~\ref{table_2}, there are two specific issues that come to mind regarding the implementation of a hybrid HetNet structure.
\begin{enumerate}
\item The spectrum in the E-band is licensed and may not be as readily available as the V-band for use in wireless and cellular networks. This presents new challenges for the deployment of hybrid HetNets by different cellular providers. However, FCC has adopted a unique licensing approach for spectrum allocation in the E-band, where the links can be quickly and economically registered over the Internet. It is also anticipated that, as the E-band is used more heavily due to spectrum needs, the licensing regulations on this band may be further relaxed.
\item The high antenna gain required for the use of the E-band spectrum may render the application of this spectrum impractical in small and portable devices. Therefore, based on the current regulations, the E-band may be mainly utilized to carry the backhaul between access points and BSs, which in any case, are long-distance links. However, as outlined above, it is anticipated that as the need for bandwidth and higher throughputs in wireless networks continue to grow, the large bandwidth in the E-band will also be made available for use via less restrictive regulations.
\end{enumerate}

%

%
%
%
%
%
%
%
%



\vspace*{-8pt}
\section{Summary}\label{sec:conc}

In order to reduce the impact of interference in HetNets and more effectively utilize the available spectrum in the millimeter-wave band, this paper proposed a novel hybrid HetNet paradigm. In a hybrid HetNet, the V-band is used to establish short range and ultra high-speed point-to-point links, e.g., within femtocells. Due to significant channel attenuation in the $60$ GHz band or the V-band, these short range links do not result in significant inter-user interfere. On the other hand, the $70$/$80$ GHz or the E-band is used to establish the links with longer ranges, e.g., to interconnect cellular BSs. The E-band can support such links, since compared to the V-band, channel attenuation in the E-band is much smaller and the regulated maximum transmit power is much greater. Our simulation results demonstrated that by employing the characteristics of both bands, a hybrid HetNet configuration can greatly enhance the throughput of millimeter-wave networks. Moreover, two competing transceiver structures for hybrid HetNets were proposed and their advantages and disadvantages, from a practical standpoint, were discussed. We also outlined the regulatory challenges that need to be addressed to make a hybrid HetNet paradigm viable and provided preliminary solutions for each issue. Finally, although in this paper, the concept of HetNets has been discussed from a physical layer point-of-view, the successful deployment of such networks requires some modification to the higher layers. For example, the use of both V- and E-bands may be determined by the higher layers based on the throughput needs of the user, while the physical layer facilitates the utilization of each band. Detailing these changes is beyond the scope of this paper and is subject of future research.

\bibliographystyle{IEEEtran}
\bibliography{IEEEabrv,Bay_PN}

\end{document}